\documentclass[11pt]{article}

\usepackage{amsmath}
\usepackage{amssymb}
\usepackage{extarrows}
\usepackage{graphicx}
 \usepackage{indentfirst}
\usepackage{hyperref}

\renewcommand{\baselinestretch}{1.3}
  \renewcommand{\arraystretch}{1.2}
\begin{document}

 \title{A note  on ``achieving security, robust cheating resistance, and
high-efficiency for outsourcing large matrix multiplication
computation to a malicious cloud"}

 \author{Zhengjun Cao$^{1}$, \quad Lihua Liu$^{2,*}$}
  \footnotetext{ $^1$Department of Mathematics, Shanghai University, Shanghai,
  China. \\
        $^2$Department of Mathematics, Shanghai Maritime University,
  China.   $^*$\,\textsf{liulh@shmtu.edu.cn} }

 \date{}\maketitle

\begin{quotation}
 \textbf{Abstract}. We show that the  Lei et al.'s scheme [Information Sciences, 280 (2014), 205-217] fails,  because the verifying equation does not hold over the infinite field $\mathbb{R}$. For the field $\mathbb{R}$, the computational errors should be considered seriously. We also remark that  the incurred communication cost in the scheme could be overtake the computational gain, which makes it somewhat artificial.

  \textbf{Keywords.}  Cloud computing, outsourcing computation,  matrix multiplication, malicious server.
 \end{quotation}

 \section{Introduction}

Recently, Lei et al. \cite{L14} have proposed a scheme for outsourcing matrix multiplication
computation over infinite field $\mathbb{R}$. In this note, we show that the  Lei et al.'s scheme fails  because the verifying equation does only hold over any finite field, instead of the infinite field $\mathbb{R}$. For the field $\mathbb{R}$, the computational errors, especially rounding errors,  should be considered carefully.
We also remark that  the incurred communication cost in the scheme could be overtake the computational gain, which makes it somewhat unpractical.

\section{Review of the Lei et al.'s  scheme}

Let $\textbf{X}(i, j), x_{i,j}$ or $x_{ij}$ denote the entry in $i$th row and $j$th column in matrix $\textbf{X}$. Define $\delta_{x, y}=1,$ if $x=y$;  $\delta_{x, y}=0,$ if $x\neq y$. Given a matrix $\textbf{X} \in \mathbb{R}^{m\times n}$ and a matrix $\textbf{Y} \in  \mathbb{R}^{n\times s}$, the resource-constrained client wants to securely outsource the computation of $\textbf{X}\textbf{Y}$ to the cloud. The scheme can be described as follows (see Table 1).

\begin{center}

Table 1: The Lei et al.'s scheme for outsourcing matrix multiplication
computation\vspace*{3mm}

\begin{tabular}{|ll|}
  \hline
  Client &    Server \\ \hline
     \emph{Setup}. Pick three sets of random numbers   &    \\
    $\{\alpha_1, \cdots,   \alpha_m\}, \
   \{\beta_1, \cdots,   \beta_n\}\
    \{\gamma_1, \cdots,   \gamma_s\}.$  &    \\
      Generate three random permutations:   &    \\
       $\pi_1$ over $\{1, \cdots, m\},$
      $\pi_2$ over  $\{1, \cdots, n\},$ &    \\
      $\pi_3$ over $\{1, \cdots, s\}.$
            Set them as the secret key.  &    \\ \hline
      \underline{\emph{Input}}  $\textbf{X} \in \mathbb{R}^{m\times n}$,  $\textbf{Y} \in  \mathbb{R}^{n\times s}$.  &   \\
      \emph{\underline{Transformation}}& \\
         \qquad $\textbf{X}'(i, j)=(\alpha_i/\beta_j)\textbf{X}(\pi_1(i), \pi_2(j))$, &    \\
     \qquad   $\textbf{Y}'(i, j)=(\beta_i/\gamma_j)\textbf{Y}(\pi_2(i), \pi_3(j))$.   & \\
       \underline{\emph{Outsourcing}}\ \  Send $\textbf{X}', \textbf{Y}'$ to the server. &  Compute $\textbf{Z}'=\textbf{X}'\textbf{Y}'$\\
       &    and return it to the client.  \\
      \underline{\emph{Composition}} &  \\
      $\textbf{Z}(i, j)=\left(\gamma_{\pi_3^{-1}(j)}/\alpha_{\pi_1^{-1}(i)}\right)\textbf{Z}'(\pi^{-1}_1(i), \pi^{-1}_3(j))$.  &   \\
     \underline{\emph{Verification}} &  \\
     Pick an $s\times 1$ random 0/1 vector $\textbf{r}$. &    \\
     Check that $\textbf{X}(\textbf{Y}\textbf{r})-\textbf{Z}\textbf{r}\stackrel{?}{=}(0, \cdots, 0)^{T}$. &    \\
     Repeat the  process $l$ times.& \\
     \underline{\emph{Output}} $\textbf{Z}$.  &    \\
   \hline
\end{tabular}\end{center}

The correctness of this procedure can be argued as follows.
  Set the matrixes $\textbf{P}_1, \textbf{P}_2, \textbf{P}_3$ as
     $$\textbf{P}_1(i, j)=\alpha_i\delta_{\pi_1(i), j}, \  \ \textbf{P}_2(i, j)=\beta_i\delta_{\pi_2(i), j},\ \  \textbf{P}_3(i, j)=\gamma_i\delta_{\pi_3(i), j}.  $$
    Then
    $$\textbf{P}_1^{-1}(i, j)=(\alpha_j)^{-1}\delta_{\pi^{-1}_1(i), j},\
    \textbf{P}_2^{-1}(i, j)=(\beta_j)^{-1}\delta_{\pi^{-1}_2(i), j}, \
    \textbf{P}_3^{-1}(i, j)=(\gamma_j)^{-1}\delta_{\pi^{-1}_3(i), j}. $$
Hence,
$$\textbf{P}_1\textbf{X}\textbf{P}_2^{-1}=(\alpha_i/\beta_j)\textbf{X}(\pi_1(i), \pi_2(j))=\textbf{X}'(i, j),$$
 $$ \textbf{P}_2\textbf{Y}\textbf{P}_3^{-1}=(\beta_i/\gamma_j)\textbf{Y}(\pi_2(i), \pi_3(j))=\textbf{Y}'(i, j),$$
$$\textbf{P}_1^{-1}\textbf{Z}'\textbf{P}_3=\left(\gamma_{\pi_3^{-1}(j)}/\alpha_{\pi_1^{-1}(i)}\right)\textbf{Z}'(\pi_1^{-1}(i), \pi_3^{-1}(j))=\textbf{Z}(i, j).$$
Since $\textbf{X}'\textbf{Y}'=\textbf{Z}'$, we have
$$\textbf{Z}=\textbf{P}_1^{-1}\textbf{X}'\textbf{Y}'\textbf{P}_3= \textbf{P}_1^{-1} \textbf{P}_1\textbf{X}\textbf{P}_2^{-1} \textbf{P}_2\textbf{Y}\textbf{P}_3^{-1}\textbf{P}_3=\textbf{X}\textbf{Y}.$$
Thus, $\textbf{X}(\textbf{Y}\textbf{r})=\textbf{Z}\textbf{r}.$

Unfortunately,  the above reasoning process is true only in some symbolic computing environments. But in a practical floating-point number system,
computational errors involved in the above procedure  should be considered seriously.

\section{The checking mechanism in the scheme fails}

\subsection{The checking equation holds only over any finite fields}

Suppose that a floating-point number system is characterized by four integers \cite{H01}: base $\chi$, precision $p$, exponent range $[L, U]$. Then its accuracy can be characterized by a quantity known as \emph{machine precision}, $\epsilon$.
If a given real number $x$ is not exactly representable as a floating-point number, then it must be approximated by some ``nearby" floating-point number. The process of choosing fl$(x)$ to approximate $x$ is called rounding, and the error introduced by such an approximation is called rounding error.

Consider the simple computation $x(y+z)$. In floating-point arithmetic we have
$$\mbox{fl}(y+z)=(y+z)(1+\theta_1), \ \ \mbox{with}\, |\theta_1|\leq \epsilon,  $$
so that
\begin{eqnarray*}
\mbox{fl}(x(y+z))&=& (x((y+z)(1+\theta_1)))(1+\theta_2), \ \mbox{with}\, |\theta_2|\leq \epsilon \\
&=& x(y+z)(1+\theta_1+\theta_2+\theta_1\theta_2)\\
&\approx &x(y+z)(1+\theta_1+\theta_2)\\
&=& x(y+z)(1+\theta), \ \mbox{with}\, |\theta|=|\theta_1+\theta_2|\leq 2\epsilon.
\end{eqnarray*}

Let $\textbf{X}=(x_{ij})_{m\times n}, \textbf{Y}=(y_{ij})_{n\times s}$,
 $\textbf{r}=(r_1, \cdots, r_s)^{T} $.  Then the first entry  of
$ \textbf{X}(\textbf{Y}\textbf{r})$ is
 \begin{eqnarray*}
 \Sigma_{j=1}^{n}x_{1j}(\Sigma_{i=1}^{s}\, y_{ji}r_{i})&=& x_{11}(y_{11}r_1+y_{12}r_2+\cdots+y_{1s}r_s) \\
 & & + x_{12}(y_{21}r_1+y_{22}r_2+\cdots+y_{2s}r_s)\\
 && +\cdots\\
 && + x_{1n}(y_{n1}r_1+y_{n2}r_2+\cdots+y_{ns}r_s)\\
 &=& r_1\Sigma_{i=1}^{n}\, x_{1i}y_{i1}+ \cdots+  r_s\Sigma_{i=1}^{n}\, x_{1i}y_{is}
 \end{eqnarray*}
Since
\begin{eqnarray*}
\textbf{Z}(1, 1)&=&\frac{\gamma_{\pi_3^{-1}(1)}}{\alpha_{\pi_1^{-1}(1)}}\textbf{Z}'(\pi_1^{-1}(1), \pi_3^{-1}(1))
  =\frac{\gamma_{\pi_3^{-1}(1)}}{\alpha_{\pi_1^{-1}(1)}}
 \left[ \Sigma_{k=1}^n X'(\pi_1^{-1}(1),k)Y'(k,\pi_3^{-1}(1))\right] \\
 &=& \frac{\gamma_{\pi_3^{-1}(1)}}{\alpha_{\pi_1^{-1}(1)}}
   \left[\left(\frac{\alpha_{\pi_1^{-1}(1)}}{\beta_1}
 x_{1,\pi_2(1)}\right)
  \left(\frac{\beta_1}{\gamma_{\pi_3^{-1}(1)}}
   y_{\pi_2(1),1}\right)+ \right.\\
   & &\cdots+ \left.   \left(\frac{\alpha_{\pi_1^{-1}(1)}}{\beta_n}
 x_{1,\pi_2(n)}\right)
  \left(\frac{\beta_n}{\gamma_{\pi_3^{-1}(1)}}
   y_{\pi_2(n),1}\right)\right]\\
  \vdots  & &
   \end{eqnarray*}
\begin{eqnarray*}
\textbf{Z}(1, s)&=&\frac{\gamma_{\pi_3^{-1}(s)}}{\alpha_{\pi_1^{-1}(1)}}\textbf{Z}'(\pi_1^{-1}(1), \pi_3^{-1}(s))
  =\frac{\gamma_{\pi_3^{-1}(s)}}{\alpha_{\pi_1^{-1}(1)}}
 \left[ \Sigma_{k=1}^n X'(\pi_1^{-1}(1),k)Y'(k,\pi_3^{-1}(s))\right] \\
 &=& \frac{\gamma_{\pi_3^{-1}(s)}}{\alpha_{\pi_1^{-1}(1)}}
   \left[\left(\frac{\alpha_{\pi_1^{-1}(1)}}{\beta_1}
 x_{1,\pi_2(1)}\right)
  \left(\frac{\beta_1}{\gamma_{\pi_3^{-1}(s)}}
   y_{\pi_2(1),1}\right)+ \right.\\
   & &\cdots+ \left.   \left(\frac{\alpha_{\pi_1^{-1}(1)}}{\beta_n}
 x_{1,\pi_2(n)}\right)
  \left(\frac{\beta_n}{\gamma_{\pi_3^{-1}(s)}}
   y_{\pi_2(n),1}\right)\right]
   \end{eqnarray*}
the first entry of $\textbf{Z}\textbf{r}$ is
\begin{eqnarray*}
& &  \frac{\gamma_{\pi_3^{-1}(1)}}{\alpha_{\pi_1^{-1}(1)}}
   \left[\left(\frac{\alpha_{\pi_1^{-1}(1)}}{\beta_1}
 x_{1,\pi_2(1)}\right)
  \left(\frac{\beta_1}{\gamma_{\pi_3^{-1}(1)}}
   y_{\pi_2(1),1}\right)+
   \cdots  +  \left(\frac{\alpha_{\pi_1^{-1}(1)}}{\beta_n}
 x_{1,\pi_2(n)}\right)
  \left(\frac{\beta_n}{\gamma_{\pi_3^{-1}(1)}}
   y_{\pi_2(n),1}\right)\right] r_1
  \\
  & & \\
  & +&  \cdots  \\
  & & \\
 & +&  \frac{\gamma_{\pi_3^{-1}(s)}}{\alpha_{\pi_1^{-1}(1)}}
   \left[\left(\frac{\alpha_{\pi_1^{-1}(1)}}{\beta_1}
 x_{1,\pi_2(1)}\right)
  \left(\frac{\beta_1}{\gamma_{\pi_3^{-1}(s)}}
   y_{\pi_2(1),1}\right)+ \right.
   \cdots \left.
     +    \left(\frac{\alpha_{\pi_1^{-1}(1)}}{\beta_n}
 x_{1,\pi_2(n)}\right)
  \left(\frac{\beta_n}{\gamma_{\pi_3^{-1}(s)}}
   y_{\pi_2(n),1}\right)\right] r_s.
   \end{eqnarray*}
 Since  $\{\alpha_1, \cdots,   \alpha_m\}, \
   \{\beta_1, \cdots,   \beta_n\}\
    \{\gamma_1, \cdots,   \gamma_s\}$ are randomly chosen in $\mathbb{R}$, the total rounding error in  the above equation  approximates to
    $\bar x_1\bar  y_1 ns\epsilon$,     where $\bar x_1=\frac{1}{n}\Sigma_{i=1}^{n} x_{1i}, \,\bar  y_1=\frac{1}{n}\Sigma_{j=1}^{n}y_{j1}$, and $\epsilon$ is the machine precision. Therefore,
   the practical computational result is
    $$ \textbf{X}(\textbf{Y}\textbf{r})- \textbf{Z}\textbf{r}=  (\bar x_1\bar  y_1 ns\epsilon, \cdots, \bar x_m\bar  y_m ns\epsilon)^T.$$
   Thus, the original checking mechanism in the Lei et al.'s  scheme fails. The authors did not pay more attentions to the  differences between the arithmetic over the infinite field $\mathbb{R}$ and that over any finite field.

   To fix the scheme, one has to check that
   $$\left|\frac{v_i}{u_i}-1\right|\leq \lambda ns \epsilon, \quad i=1, \cdots, m$$
    where
    $$\textbf{X}(\textbf{Y}\textbf{r})=(u_1, \cdots, u_m)^T, \quad \textbf{Z}\textbf{r}=(v_1, \cdots, v_m)^T,$$
     and $\lambda$ is a fault-tolerant parameter. If all $m$ inequalities are true, then output $\textbf{Z}$.

    \subsection{The revised version of Lei et al.'s scheme is insecure}

    Though the computational errors have been considered in the revised version of Lei et al.'s scheme, we show it is insecure because the malicious server can cheat the client to accept a wrong result.

    To do this, the malicious server computes $\textbf{Z}'=\textbf{X}'\textbf{Y}'$ and returns $\widehat{\textbf{Z}'}$ to the client, where
    $$\widehat{\textbf{Z}'}=\textbf{Z}'+\left(
                                      \begin{array}{cccc}
                                        \rho & 0& \cdots & 0 \\
                                        0&0 &  \cdots & 0 \\
                                        \vdots&\vdots &\vdots &\vdots \\
                                        0&0 & \cdots & 0 \\
                                      \end{array}
                                    \right),
    $$
    $\rho=(\chi-1)\epsilon$,  $ \chi$ is the base of the underlying floating-point number system.     In such case, we have
     $$\textbf{Z}(\pi_1(1), \pi_3(1))=\left(\gamma_{1}/\alpha_{1}\right)[\textbf{Z}'(1, 1)+\rho],$$
      $$ \textbf{Z}\textbf{r}=(v_1, \cdots, v_{\pi_1(1)}+\frac{\gamma_{1}}{\alpha_{1}} r_{\pi_3(1)}\rho,\cdots, v_m)^T $$
      Hence, $$\left|\frac{v_{\pi_1(1)}+\frac{\gamma_{1}}{\alpha_{1}} r_{\pi_3(1)}\rho}{u_{\pi_1(1)}}-1\right|\leq \left|\frac{v_{\pi_1(1)}}{u_{\pi_1(1)}}-1\right|+\left|\frac{\frac{\gamma_{1}}{\alpha_{1}}\rho}{u_{\pi_1(1)}}\right|.  $$
      The probability of the event that
      $$\left|\frac{v_{\pi_1(1)}}{u_{\pi_1(1)}}-1\right|+\left|\frac{\frac{\gamma_{1}}{\alpha_{1}}}{u_{\pi_1(1)}}\right|\cdot (\chi-1)\epsilon \leq \lambda ns \epsilon $$ approximates to 1, because $n, s $ are supposed to be sufficiently large. Thus, the malicious server can cheat the client to accept
      a wrong result $\textbf{Z}$, where
      $$\textbf{Z}(i, j)=\left(\gamma_{\pi_3^{-1}(j)}/\alpha_{\pi_1^{-1}(i)}\right)\widehat{\textbf{Z}'}(\pi^{-1}_1(i), \pi^{-1}_3(j)).$$

    \section{Further discussions}

 The authors \cite{L14} claim that  the proposed scheme is feasible due to there exists a gap
between $O(mn + ns + ms)$ and $O(mns)$. Therefore, as long as $m, n, s$ become sufficiently large, the scheme allows the client to outsource the computations to the cloud and gain substantial computational savings.

We now want to stress that the client  has to interact with the untrusted cloud via a non authenticated link, by sending $\textbf{X}', \textbf{Y}'$ and receiving $\textbf{Z}'$. From the practical point of view,
  the communication costs (including authentication of the exchanged data, the possible underlying encryption/decryption, the time delay during the interaction, etc.) could be far more than the above computational gain. That means the scheme is somewhat unrealistic.

One might suggest  to constrain the Lei et al.'s scheme to some finite fields.  But we would like to stress  that matrix multiplication
computation over finite fields is of little practical importance. For example, the most popular linear programming cannot be constrained to any finite field.

\section{Conclusion}
We show that the Lei et al.'s scheme for outsourcing matrix multiplication computation over the infinite field $\mathbb{R}$ is insecure. We think, the problem that what computations are worth delegating privately by individuals and companies to an untrusted cloud remains open. The cloud computing community has not yet found a good for-profit model convincing individuals to pay for this or that computational service.

\end{document}